\documentclass[
hf,
]{ceurart}

\usepackage{cleveref}
\usepackage[vlined, linesnumbered ]{algorithm2e}

\usepackage{amsmath,amsfonts,amssymb,mathrsfs}
\begin{document}


\conference{ArXiv preprint}

\title{InsightiGen: a versatile tool to generate insight for an academic systematic literature review}

\author[1]{Ardeshir Shojaeinasab}[%
email=ardeshir@uvic.ca
]

\author[1,2]{Masoud Jalayer}[%
orcid=0000-0001-8013-8613,
email=masoudjalayer@uvic.ca,
]

\author[1,2]{Homayoun Najjaran}[%
email=najjaran@uvic.ca,
]

\address[1]{Department of Electrical and Computer Engineering, University of Victoria, Victoria, BC,  V8P 5C2, Canada}

\address[2]{Department of Mechanical Engineering, University of Victoria, Victoria, BC,  V8P 5C2, Canada}

\maketitle
\begin{abstract}
A comprehensive literature review has always been an essential first step of every meaningful research. In recent years, however, the availability of a vast amount of information in both open-access and subscription-based literature in every field has made it difficult, if not impossible, to be certain about the comprehensiveness of one’s survey. This subsequently can lead to reviewers' questioning of the novelties of the research directions proposed, regardless of the quality of the actual work presented. In this situation, statistics derived from the published literature data can provide valuable quantitative and visual information about research trends, knowledge gaps, and research networks and hubs in different fields. Our tool provides an automatic and rapid way of generating insight for systematic reviews in any research area.
\end{abstract}

\section{INTRODUCTION} \label{introduction}

A systematic literature review must contain some statistics about the literature and the corresponding insights for the researchers who want to pursue the domain. Reading a systematic review that contains valuable statistics informs readers about the influential clusters, institutes and authors in a specific area. It prevents the readers missing seminal research in a particular domain and keeps them updated about the critical papers of the area.

In 2017, Xiao et al. outlined the steps of writing a comprehensive and systematic literature review \cite{AA1}. The steps include: formulate the problem, develop and validate the review protocol, search the literature, screen for inclusion, and finally asses the quality acting as a fine sieve to refine the full-text articles and preparing the pool of studies for data extraction and synthesis. In this process, the last two steps serve the purpose of extracting, analyzing and synthesizing data.

Several tools were developed to address researchers' needs regarding literature review and bibliography analysis. The Connected Papers project is a website for exploring connected papers to a particular paper in a visual graph. It opens the ability to discover important recent papers visually and makes researchers independent of keeping lists. Hence, It helps to ensure you have not missed an important paper. On top of that, this tool has another option to find the most relevant prior and derivative works. It is possible to find notable ancestral works in any topic of study using the prior works feature. By using the derivative works view, researchers can identify literature reviews on the subject of the fed paper, as well as recently published state-of-the-art that followed one's input paper\cite{connectedpapers}. CoCites searches for articles that are regularly cited alongside articles of relevance. As a result, it can be used in research studies when the purpose is to identify comparable articles. Additionally, because the search results are given with the articles' citation counts, it can be used to determine the most well-known or key articles on niche topics. This is helpful information for conducting research on areas outside of one's area of expertise \cite{cocites}. One of the other tools which are more concentrated on text mining is JSTOR. This tool helps create the reading list in a more organized and easy way. This tool has a feature to search for similar and co-cited papers related to the paper researcher requests. Also, it has a library option to add any of resulted papers to the researcher's library \cite{jstor}. CSAuthors is a tool that is developed to output information about co-authorship. This tool developed a distance calculator which indicates the shortest path length between two researchers regarding the co-authorship using the Dijkstra algorithm \cite{csauthors}. Litmaps is developed to give this opportunity to authors to make a visualized citation graph from a batch of papers by importing them into the tool \cite{litmaps}. One of the most powerful literature review tools that have been developed is VOSviewer by Leiden University researchers. VOSviewer is software that allows researchers to create and visualize bibliometric networks. These networks can be built via citation, bibliographic coupling, co-citation, or co-authorship relationships, and they can include journals, researchers, or individual articles. Text mining functionality is also available in VOSviewer, which may be used to create and visualize co-occurrence networks of relevant terms retrieved from a corpus of scientific literature \cite{vosviewer}.

The authors developed the InsightiGen and used it to collect and analyze the literature data for the recently published systematic literature review in the intelligent manufacturing execution system \cite{BB1}. InsightiGen is an open-source tool that is available through a GitHub repository\footnote{\href{https://github.com/Ardeshir-Shon/systematicReviewGenerator}{InsightiGen GitHub Repository}}. The tool primarily automates the last two steps of the systematic literature review that are extracting data and generating insight from the extracted data. The tool performs data cleaning, data processing, graph analysis and natural language processing to generate and visualize meaningful information.

The process is elaborated in the following sections.

\section{METHODOLOGY} \label{sec:method}
\begin{figure}[h]
    \centering
    \includegraphics[width=12cm]{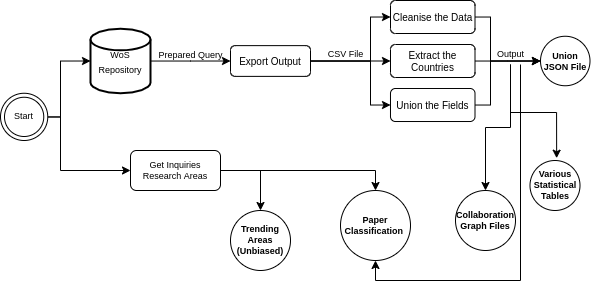}
    \caption{An overview of the tool in a higher abstraction}
    \label{fig:flow}
\end{figure}

The paper proposes a tool for automating the last two aforementioned systematic review steps. Once the users outlined the research questions and identified the keyword queries, they can employ Insightigen. In order to get a better result, it is possible to define a multi-level keyword query, separately for the research domain and for the research methods, as the authors have done in \cite{BB1}.
In what follows, the authors elaborate more on some insights that the tool yields. The tool is designed and developed such that it can be used regardless of the area. It is configurable and researchers based on their needs can use the tool, even for sentiment analysis and extracting the trending topic. 
Figure \ref{fig:flow} presents a brief overview of the tool.

\subsection{Parsing \& Cleaning the Data}
At the very first step, there is a need to clean and parse the data. This is because the exported data is not clean and in some data fields, useful data are implicit. For instance, no field indicates the nationality of the affiliations. Also, the collaborations of the affiliations are not defined separately. In addition, there are a lot of missing values in the exported data that should be considered. InsightiGen produces a JSON file containingl data in a well-structured, cleaned and human-readable format thathateasy to import and work on. The JSON file has all fields such as title, DOI, abstract, author names, author affiliations, affiliations' countries and keywords.
The JSON file is easy to import in any tool for further analysis and there is no need for data cleaning and parsing anymore.

\subsection{Extracting the Trending Topics}
To determine the trending topics, this paper uses a new method adopted from a popular classic technique in natural language processing (NLP) called TF-IDF. The main idea behind TF-IDF is to determine the frequency of words as well as the commonness of each word \cite{CC1} \cite{BB2}. This paper uses the same idea for calculating commonness in the domain of trending topics. Grouped based on each topic in a particular year, we calculate what portion of the papers in the last three years have been related to the same topic. A higher percentage means that the topic was commonly used in the previous three years; hence it is more common. On the other hand, if the percentage is low, the topic in this area, although not common, it is more niche. Specifically, equation \ref{trendiness}, where $\rho$ denotes the number of papers that use a certain topic in the year and $\delta$ denotes the number of papers that have used certain topic in previous three years and $N$ is number of all papers in the area in previous three years, is determining trendiness.

\begin{equation}\label{trendiness}
    trendiness = \frac{\rho}{\log_2  \frac{\delta}{N}}
\end{equation}

\subsection{Influential Component In a Graph} \label{sec:graph_centrality}

Identifying influential nodes of a graph is a well-known network science problem with many important applications in other disciplines such as social sciences \cite{DD1}, neuroscience \cite{EE1}, finance \cite{FF1}, and economics \cite{AA2}. Network science is an emerging interdisciplinary field of study that can describe a variety of systems with graph structures \cite{AA3}.

The influence of a node is chiefly impacted by the topology of the graph to which it belongs. In fact, the majority of the proposed models that identify influential nodes only the structural information, which allows the wide applications independent to the specific dynamical processes under consideration. The concept centrality was mainly proposed to answer the question that how to characterize a node’s importance according to the network topology \cite{AA4}. A systematic literature review should include the information about the influential authors, institutions and countries of the research area.

A graph with a vector of $V$ that stand for the nodes and a matrix of $E$ that represents the connectivity between each pair of nodes, can be mathematically defined as $G=(V,E)$; where $n=|V|$ and $u=|E|$ denote the numbers of elements in $V$ and $E$, respectively.

InsightiGen outputs three networks: (i) co-authorship, (ii) institutional collaborations, and (iii) country-wise collaborations; in which the nodes represent the authors, the affiliations and the countries, respectively. The edges of each graph show the appearance of each pair of nodes in a paper. And the weights are the number of papers at each edge. All the resulted networks are weighted and undirected.

\begin{figure}
    \centering
    \includegraphics[width=8cm]{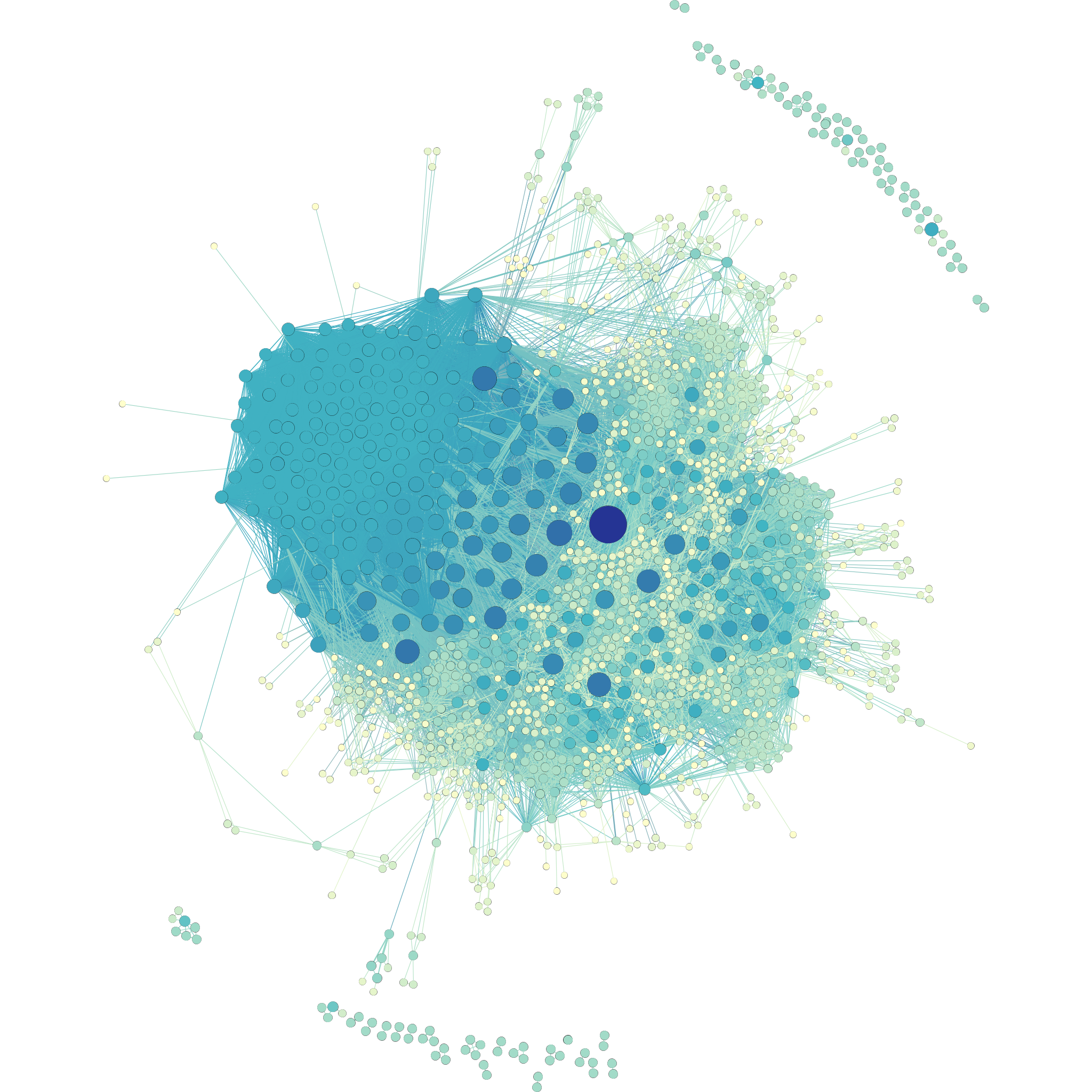}
    \caption{The output network structure of institutional collaborations for a medical-related random query - The node sizes are proportionate to their PageRank values}
    \label{fig:institutional1}

    \centering
    \includegraphics[width=8cm]{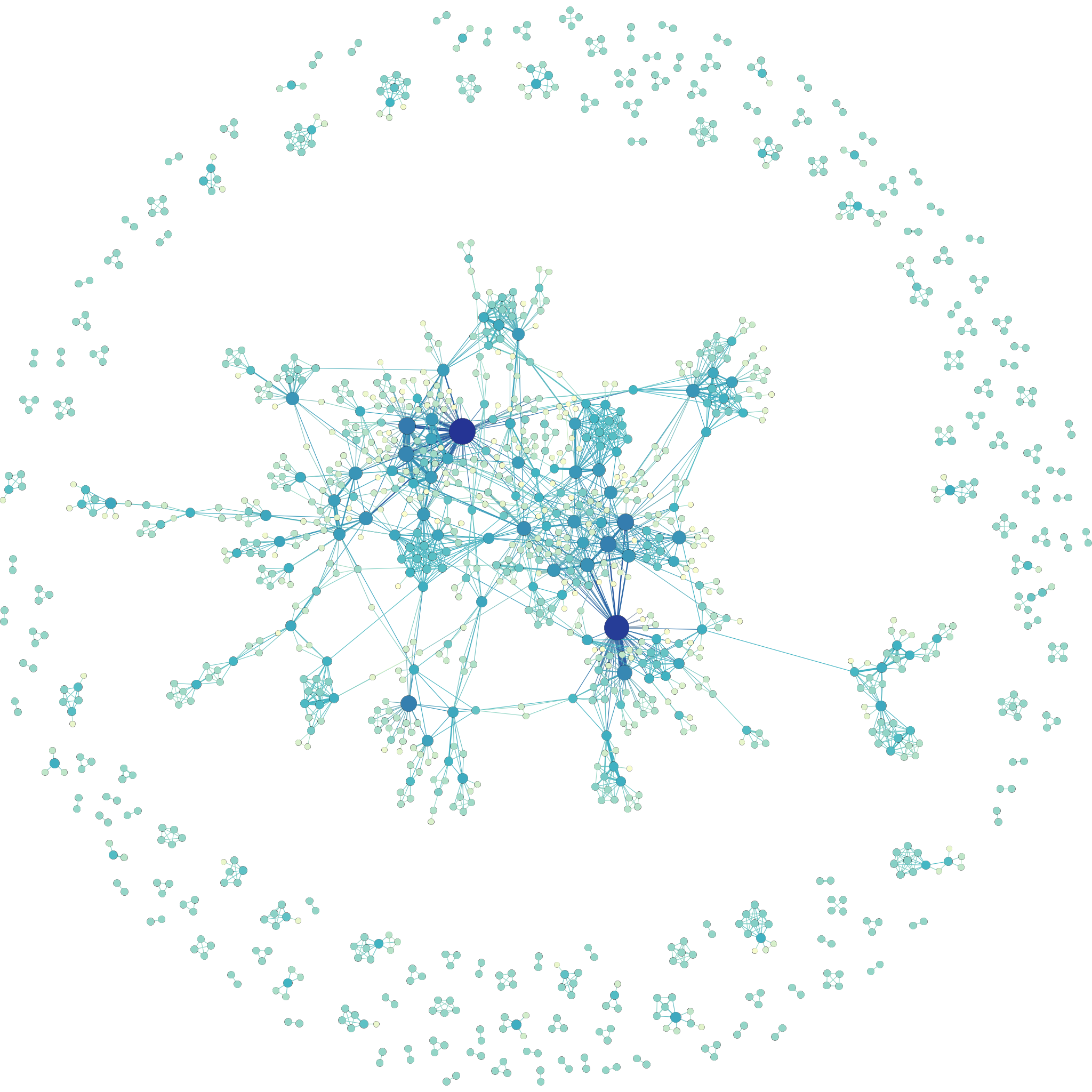}
    \caption{The output network structure of institutional collaborations for a astrophysics-related random query - The node sizes are proportionate to their PageRank values}
    \label{fig:institutional2}
\end{figure}

\begin{figure}[h]
    \centering
    \includegraphics[width=\linewidth]{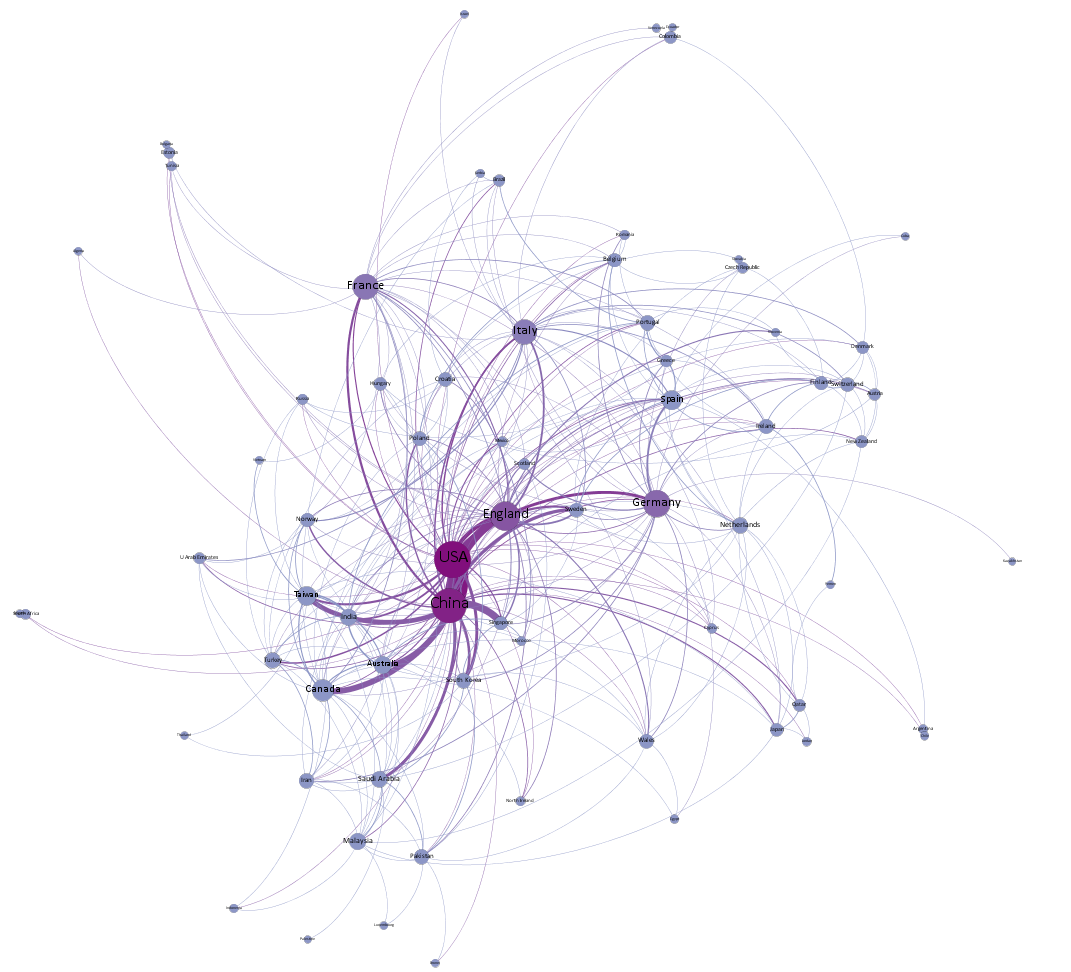}
    \caption{The output network structure of country collaborations for a random Web of Science query - The node sizes are proportionate to their Betweenness values}
    \label{fig:countriesGraph}
\end{figure}

Figures~\ref{fig:institutional1} and ~\ref{fig:countriesGraph} illustrate the output networks of InsightiGen corresponding to two random queries for institutional collaborations and country-wise collaborations, respectively. Topological analysis of the different research areas is another factor that can be inferred from the generated collaboration graphs. For instance, generated graphs in Figures~\ref{fig:institutional1} and ~\ref{fig:institutional2} imply that institutions in the medical area are tightly connected. While, for the astrophysics, institutions are more isolated and there is less collaboration. 

Betweenness centrality was firstly proposed by Bavelas in 1948 \cite{AA7} to measure the node potential powers in controlling the information flow in a network. In 1977, Freeman \cite{AA8} generalized the graph theoretical notion of this centrality and extended it to both connected and unconnected networks, showing it the way it is used today. Generally, there exists more than one shortest paths starting from node $v_s$ and ending at $v_t$. The controllability of information flow of $v_i$ can be computed by counting all of the shortest paths passing through $v_i$. Thus, the betweenness centrality of node $v_i$ can be defined as
\begin{equation}
    BC(i)=\sum_{i\neq s, i\neq t, s \neq t}\frac{g^{i}_{st}}{g_{st}}
\end{equation}

where $g_{st}$ is the number of the shortest paths between $v_s$ and $v_t$ and $g^{i}_{st}$ is the number of paths which pass through $v_i$ among all the shortest paths between $v_s$ and $v_t$. It can be proven that in a network with $n$ nodes. the maximum possible betweenness centrality value of a node is $(n-1)(n-2)/2$. Hence, the normalized betweenness centrality of node $v_i$ will be:

\begin{equation}
    BC(i)= \frac{2}{(n-1)(n-2)} \sum_{i\neq s, i\neq t, s \neq t}\frac{g^{i}_{st}}{g_{st}}
\end{equation}

PageRank algorithm \cite{AA5} is another renowned centrality measurements that is initially introduced to rank websites in Google search engine and other commercial scenarios \cite{AA6}. PageRank measures the importance of each website by random walking on the network constructed from the relationships of web pages. Similar to eigenvector centrality, PageRank supposes that the importance of a web page is determined by both the quantity and the quality of the pages linked to it. Initially, each node (i.e., page) gets one unit PR value. Then every node evenly distributes the PR value to its neighbors along its outgoing links. PR value for node $v_i$ at $t$ step can be mathematically defined as follows:

\begin{equation}
    PR_{i}(t)=\sum^{n}_{j=1} a_{ji} \frac{PR_{j}(t-1)}{k_{j}^{out}}
\end{equation}


\subsection{Semantic Analysis Tool}
The measure of term specificity first proposed in 1972 by Karen Spärck Jones later became known as inverse document frequency or IDF. The paper Karen Spärck Jones published called “A statistical interpretation of term specificity and its application in retrieval” \cite{BB2}. Based on the term frequency normalization method proposed, IDF and some other methods for document sorting and scoring have been proposed. The Okapi BM25 is one of the methods based on the TF-IDF, widely used in the current search engines for document scoring and ranking based on a query vector. \cite{BB3}
Equation \ref{okapiBM25} demonstrates how the BM25 is able to score the relevance of a query(in this terminology, keyword vector of an area) and a document(in this terminology means a paper). In the \ref{okapiBM25} equation, $f(q_{i},D)$ is $q_{i}$'s term frequency in the document $D$, $|D|$ is the length of the document $D$ in words, and $avgdl$ is the average document length in the text collection from which documents are drawn. $k_{1}$ and $b$ are free parameters.

\begin{equation}\label{okapiBM25}
\begin{split}
score(D,Q)=\sum_{i=1}^{n}(IDF(q_{i})\cdot\frac{f(q_{i})\cdot(k_{1}+1)}{f(q_{i},D)+k_{1}\cdot(1-b+b\cdot\frac{|D|}{avgdl})})
\end{split}
\end{equation}

Another module of the InsightiGen tool is the ability to identify the semantics of the paper considering the title, keywords and abstract. To this end, the tool is implemented a Okapi BM25 algorithm to rank papers among the input keyword vectors. Each input keyword vector demonstrates a specific topic that the client passed to the tool to know the correlation of the research and that topic vector.

\section{IMPLEMENTATION \& DISCUSSION}
In this section, one example of the tool implementation is discussed as a case study to elaborate more on what insights can be driven from InsightiGen. 

As an instance, one may be interested in drawing some insights about the Smart Manufacturing Execution Systems. Let the following query (see Table~\ref{tab:include}) be the desired scope of the literature review including the keywords on the domains and technologies used in this area.

\begin{table}[ht]
\centering
\caption{The proposed two-level keyword assembly structure}
\label{tab:include}
\begin{tabular}{p{2cm}p{8cm}p{4cm}}
\toprule
Context        & Query                                                                                                                                                                                                                               & Searching field \\
\midrule
Domain & "Manufacturing Execution" or "Production Line*" or "Smart Factory" or "Manufacturing System*" or "Smart Manufacturing" or "Intelligent Manufacturing"                                                                                                                                           & Title \& Keywords \& Abstract           \\
Model / Technology   & "Computer Vision" or "Reinforcement Learning" or "Virtual Reality" or "Augmented Reality" or "Digital Twin" or "Deep Learning" or "Machine Learning" or "Machine Vision" or "Autoencoder*" or "Convolution* Net*" or "Long Short  Term" or "Blockchain" or "5G" & Title \& Keywords \& Abstract   \\
\bottomrule
\end{tabular}
\end{table}

As it is shown in Figure~\ref{fig:flow}A \color{black}, after defining the query, we can feed InsightiGen with the corresponding CSV file extracted from Web of Science database. At first, the tool can extract some statistical and chronological information about the publishers, journals, conferences, authors and their affiliations. Table~\ref{tab:journals} shows the top 20 sources based on the query, and the distribution of the publications among them, between 2009 and 2021.

\begin{table}[]
\centering
\caption{The most active journals in intelligent MES since 2009}
\label{tab:journals}
\resizebox{1\columnwidth}{!}{
\begin{tabular}{p{6cm}p{0.5cm}p{0.5cm}p{0.5cm}p{0.5cm}p{0.5cm}p{0.5cm}p{0.5cm}p{0.5cm}p{0.5cm}p{0.5cm}p{0.5cm}p{0.5cm}p{0.5cm}p{0.75cm}}
\toprule
                                                           & 2009 & 2010 & 2011 & 2012 & 2013 & 2014 & 2015 & 2016 & 2017 & 2018 & 2019 & 2020 & 2021 & \textbf{Total} \\
\midrule
IEEE Access & - & - & - & - & - & - & - & 1 & 5 & 12 & 14 & 35 & 30 & \textbf{97} \\
        Journal of Manufacturing Systems & - & - & - & - & - & - & - & - & 1 & 10 & 1 & 20 & 36 & \textbf{68} \\
        International Journal of Advanced Manufacturing Technology & - & 1 & 1 & - & - & 1 & 1 & 1 & 2 & 5 & 4 & 18 & 18 & \textbf{52} \\
        Sensors & - & - & - & - & - & - & - & - & 1 & 4 & 7 & 12 & 21 & \textbf{45} \\
        Applied Sciences-Basel & - & - & - & - & - & - & - & - & 1 & 1 & 3 & 9 & 30 & \textbf{44} \\
        IEEE Transaction on Industrial Informatics & - & - & - & - & - & 1 & - & - & 1 & 1 & 8 & 8 & 12 & \textbf{31} \\
        International Journal of Production Research & - & 3 & 2 & 2 & - & - & - & 1 & - & 1 & 4 & 11 & 7 & \textbf{31} \\
        Robotics and Computer-Integrated Manufacturing & 2 & - & 2 & - & - & - & 1 & 1 & - & 1 & 3 & 12 & 8 & \textbf{30} \\
        Journal of Intelligent Manufacturing & - & - & - & 1 & - & - & - & 1 & 1 & 1 & 3 & 11 & 8 & \textbf{26} \\
        International Journal of Computer Integrated Manufacturing & - & - & - & - & 1 & - & 2 & - & - & - & 5 & 4 & 13 & \textbf{25} \\
        Computers \& Industrial Engineering & 1 & - & 2 & - & - & - & - & 1 & 3 & 2 & 7 & 5 & \textbf{21} \\
        Computers in Industry & - & - & - & - & - & - & - & - & - & 1 & 5 & 6 & 6 & \textbf{18} \\
        Sustainability & - & - & - & - & - & - & - & - & - & - & 2 & 8 & 8 & \textbf{18} \\
        CIRP Annals - Manufacturing Technology & - & - & - & - & - & - & - & - & 1 & 1 & 3 & 4 & 5 & \textbf{14} \\
        Expert Systems With Applications & 2 & - & - & 3 & - & - & - & - & - & - & - & 1 & 5 & \textbf{11} \\
\bottomrule
\end{tabular}}
\end{table}

\begin{table}[ht]
\centering
\caption{Top influencing authors by number of papers}
\label{tab:authPapers}
\begin{tabular}{lll}
\toprule
\textbf{Author Name}    & \textbf{\#Papers} \\
\midrule
Gao, Robert X.                   & 17                 \\
Liu, Qiang                  & 14                 \\
Tao, Fei              & 14                 \\
Leng, Jiewu                    & 12                 \\
Li, Xinyu                  & 11                 \\
Gao, Liang                    & 11                 \\
Chang, Qing                   & 8                 \\
Chen, Xin                  & 8                 \\
Shiue, Yeou-Ren                   & 8                 \\
Zhang, Ding                & 7                 \\
Park, Kyu Tae               & 7                 \\
Xu, Ke               & 7                 \\
Wang, Lihui             & 6                 \\
Zhong, Ray Y.                 & 6                 \\
Nee, A. Y. C. & 6                 \\
NeeZhang, Jianjing & 6                 \\
Noh, Sang Do & 6                 \\
Zheng, Pai & 6                 \\
Wen, Long & 6                 \\
Wu, Dazhong           & 6                 \\
\bottomrule
\end{tabular}
\end{table}

\begin{figure}[h]
    \centering
    \includegraphics[width=12cm]{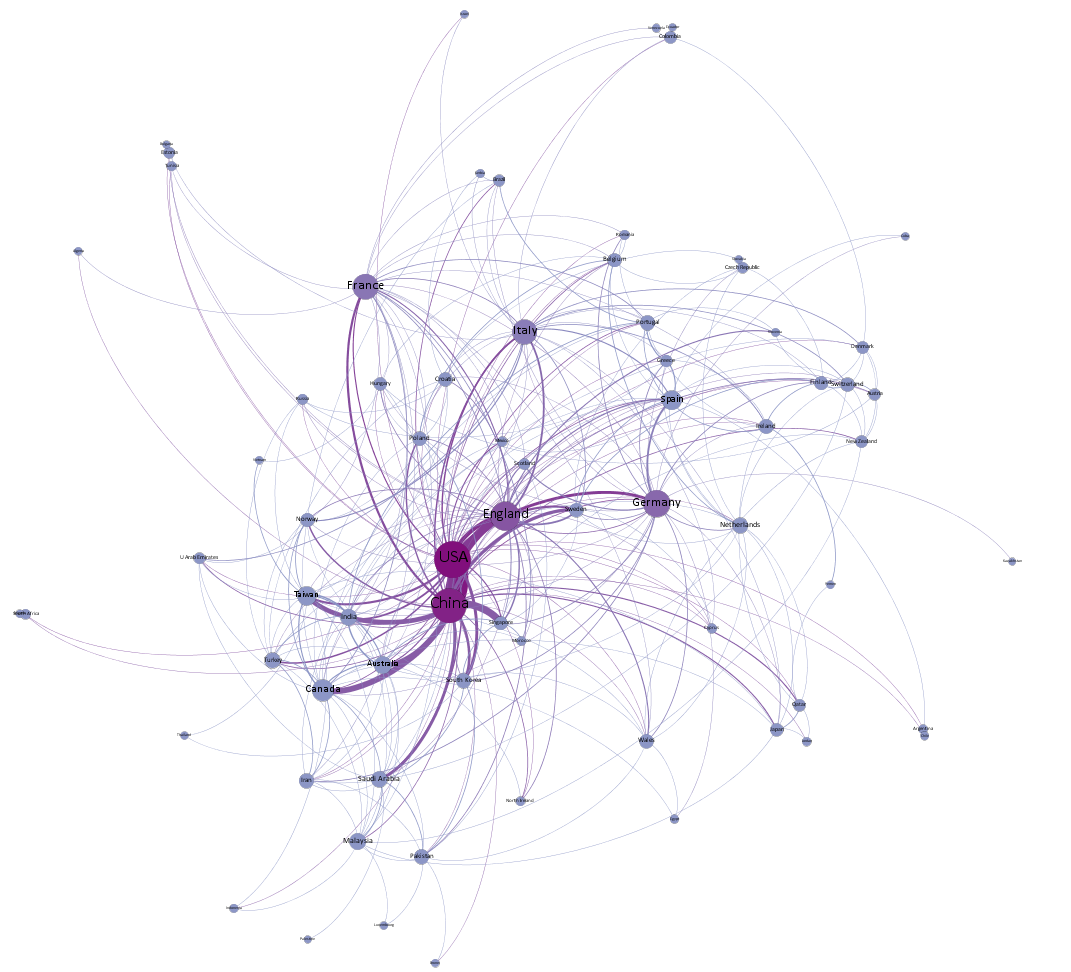}
    \caption{The output graph of country collaborations of the MES query - The node sizes are proportionate to their Betweenness values}
    \label{fig:mesGraph}
\end{figure}

To recognize the hubs of research and focus more on the research works done by the leading scholars, some might be interested in finding the most influential authors in the literature, the collaborations among the research institutes and the clusters of countries working with each other on the desired field of study. To this end, InsightiGen processes the relational data and outputs several tables and graphs as shown in Tables~\ref{tab:authPapers} to \ref{tab:authCitations} and Figure~\ref{fig:mesGraph} with the centrality measurements, as defined in Section~\ref{sec:graph_centrality}.

\begin{table}[ht]
\centering
\caption{Top influencing authors by number of citations}
\label{tab:authCitations}
\begin{tabular}{lll}
\toprule
\textbf{Author Name}    & \textbf{\#Citations} \\
\midrule
Gao, Robert X.                & 1945               \\
Tao, Fei              & 1449               \\
Yan, Ruqiang                   & 1216               \\
Gao, Liang                  & 1143               \\
Li, Xinyu                  & 1118               \\
Zhao, Rui              & 1117               \\
Wen, Long                   & 1100               \\
Mao, Kezhi                  & 1068               \\
Wang, Peng                & 993               \\
Wang, Jinjiang                   & 891               \\
Wu, Dazhong                  & 800               \\
Chen, Zhenghua                 & 780               \\
Liu, Qiang                   & 671               \\
Leng, Jiewu                  & 648               \\
Zhang, Yuyan & 639               \\
\bottomrule
\end{tabular}
\end{table}

\begin{table}[ht]
\centering
\caption{Top influencing affiliations by number of researches}
\label{tab:affPapers}
\begin{tabular}{lll}
\toprule
\textbf{Affiliation Name}    & \textbf{\#Papers} \\
\midrule
Huazhong Univ Sci \& Technol & 37 \\
        Beihang Univ & 28 \\
        Guangdong Univ Technol & 23 \\
        Xi An Jiao Tong Univ & 21 \\
        Nanyang Technol Univ & 21 \\
        Shanghai Jiao Tong Univ & 17 \\
        Zhejiang Univ & 16 \\
        Case Western Reserve Univ & 16 \\
        Chinese Acad Sci & 16 \\
        Natl Cheng Kung Univ & 15 \\
        Tsinghua Univ & 14 \\
        Univ Hong Kong & 14 \\
        Northwestern Polytech Univ & 14 \\
        Natl Tsing Hua Univ & 14 \\
        Nanjing Univ Aeronaut \& Astronaut & 13 \\
        Northeastern Univ & 12 \\
        Hong Kong Polytech Univ & 12 \\
        Univ Michigan & 12 \\
        Natl Univ Singapore & 12 \\
        Donghua Univ & 11 \\
        Sungkyunkwan Univ & 11 \\
\bottomrule
\end{tabular}
\end{table}

\begin{table}[h]
\centering
\caption{Top influencing affiliations by number of citations}
\label{tab:affCitations}
\begin{tabular}{lll}
\toprule
\textbf{Affiliation Name}    & \textbf{\#Citations} \\
\midrule
Case Western Reserve Univ & 1945 \\
        Beihang Univ & 1699 \\
        Nanyang Technol Univ & 1443 \\
        Huazhong Univ Sci \& Technol & 1384 \\
        Xi An Jiao Tong Univ & 1320 \\
        China Univ Petr & 892 \\
        Guangdong Univ Technol & 865 \\
        Natl Univ Singapore & 687 \\
        Univ Cent Florida & 490 \\
        Natl Cheng Kung Univ & 436 \\
        Grenoble Inst Technol Grenoble INP & 417 \\
        Univ Fed Rio Grande do Sul & 417 \\
        Univ A Coruna & 402 \\
        Penn State Univ & 395 \\
        Shanghai Jiao Tong Univ & 391 \\
        Univ Bremen & 373 \\
        Univ Auckland & 355 \\
        West Virginia Univ & 353 \\
        Hong Kong Polytech Univ & 352 \\
\bottomrule
\end{tabular}
\end{table}

By refining the information that InsightiGen extracts from the output of Web of Science, it automatically generates the current and previous trends in the literature. This allows the users to conclude what the research gaps are and what the future of the field under investigation will be. An example of this output is given in Table~\ref{tab:trendingTopicsPerYear}. Moreover, the tool calculates the relevance scores to a set of given topics, for each paper. Users can identify more relevant papers with this information if they want to highlight them in their survey or select a group of the most similar papers to read their full contents. Table~\ref{tab:areas} shows the relevance scores for each paper related to each requested area for an arbitrary set of papers.

\begin{table}[h]
\centering
\caption{Corresponding score for each paper regarding each area (random selected papers)}
\label{tab:areas}
\begin{tabular}{c c c c c c c }
\hline
\textbf{} & \textbf{Machine Learning} & \textbf{Blockchain} & \textbf{5G} & \textbf{Digital Twin} & \textbf{Computer Vision} & \textbf{RL} \\ \hline
Paper \#1 & 3.66 (1)                  & 0                   & 0           & 0                     & 1.34 (2)                 & 0                               \\ 
Paper \#2 & 1.78 (2)                  & 0                   & 0           & 0                     & 0.56 (3)                 & 4.8 (1)                         \\ 
Paper \#3 & 2.43 (2)                  & 2.89 (1)            & 0           & 0                     & 0.77 (3)                 & 0                               \\ \hline
\end{tabular}
\end{table}

\begin{table}[h]
\centering
\caption{The top four trending topics in the literature of smart manufacturing execution systems}
\label{tab:trendingTopicsPerYear}
\resizebox{\columnwidth}{!}{%
\begin{tabular}{ccccc}
\toprule
\textbf{Year} & \textbf{First Trending Topic}           & \textbf{Second Trending Topic}          & \textbf{Third Trending Topic} & \textbf{Fourth Trending Topic}                    \\
\midrule
\textbf{2021} & Digital Twin & Deep Learning & Reinforcement Learning & 5G \\
        \textbf{2020} & Digital Twin & Deep Learning & Blockchain & Reinforcement Learning \\
        \textbf{2019} & Digital Twin & Blockchain & Deep Learning & Computer Vision \\
        \textbf{2018} & Deep Learning & 5G & Reinforcement Learning & Digital Twin \\
        \textbf{2017} & Deep Learning & Digital Twin & Machine Learning & Blockchain \\
        \textbf{2016} & Augmented Reality & Machine Learning & Virtual Reality & Computer Vision \\
        \textbf{2015} & Augmented Reality & Virtual Reality & Computer Vision & Machine Learning \\
        \textbf{2014} & Augmented Reality & Computer Vision & Virtual Reality & Meta-Heuristic Algorithms \\
        \textbf{2013} & Deep Learning & Computer Vision & Machine Learning & Reinforcement Learning \\
        \textbf{2012} & Reinforcement Learning & Machine Learning & Meta-Heuristic Algorithms & Virtual Reality \\
        \textbf{2011} & Computer Vision & Machine Learning & Meta-Heuristic Algorithms & Virtual Reality \\
        \textbf{2010} & Computer Vision & Reinforcement Learning & Virtual Reality & Meta-Heuristic Algorithms \\
        \textbf{2009} & Meta-Heuristic Algorithms & Virtual Reality & Machine Learning & - \\
        \textbf{2008} & Machine Learning & Virtual Reality & - & - \\
\bottomrule
\end{tabular}%
}
\end{table}


\subsection{Tools Comparision}
In section\ref{introduction}, the manuscript discussed similar tools which are used in various literature reviews. Then the methodology of InsightiGen is shown in detail and a case study of the tool's usage is given.
Hence, it is crucial to have a discussion about the differences between InsightiGen and other literature review facilitator tools. At Table~\ref{tab:comparision} different systematic literature review tools are compared regarding thirteen different factors.
\begin{table}[]
\centering
\caption{Features of InsightiGen and Other Similar Tools}
\label{tab:comparision}
\resizebox{\columnwidth}{!}{%
\begin{tabular}{cccccccccccccc}
\hline
\textbf{}            & \textbf{\rotatebox{90}{Free}} & \textbf{\rotatebox{90}{Open-Source }} & \textbf{\rotatebox{90}{Web Based }} & \textbf{\rotatebox{90}{Input }}                                                                 & \textbf{\rotatebox{90}{In-House Database  }} & \textbf{\rotatebox{90}{\begin{tabular}[c]{@{}c@{}}Intermediate \\ Generated Output\end{tabular}}} & \textbf{\rotatebox{90}{Bookmark Library }} & \textbf{\rotatebox{90}{\begin{tabular}[c]{@{}c@{}}Similar Paper\\  Suggestion\end{tabular}}} & \textbf{\rotatebox{90}{Graph Generation }} & \textbf{\rotatebox{90}{\begin{tabular}[c]{@{}c@{}}Author's Related\\  Distance\end{tabular}}} & \textbf{\rotatebox{90}{\begin{tabular}[c]{@{}c@{}}Institution/Authors/ \\ Countries/ Journas\\  Statistics \end{tabular}}} & \textbf{\rotatebox{90}{Trending Areas }} & \textbf{\rotatebox{90}{\begin{tabular}[c]{@{}c@{}}Text Analyzer \\ Tool for Scoring\end{tabular}}} \\ \hline
Connected Papers     & Yes           & No                   & Yes                & Single Paper                                                                   & Yes                        & No                                                                                & Yes                                                                  & Yes                                                                          & Yes                       & No                                                                            & No                                                                                                       & No                      & No                                                                                    \\ 
CoCites              & Yes           & No                   & Yes                & Query                                                                          & Yes                        & No                                                                                & No                                                                   & Yes                                                                          & Yes                       & No                                                                            & No                                                                                                       & No                      & No                                                                                    \\ 
JSTOR                & Yes           & No                   & Yes                & Single Paper                                                                   & No                         & No                                                                                & Yes                                                                  & Yes                                                                          & No                        & No                                                                            & No                                                                                                       & No                      & Yes                                                                                   \\ 
VOSViewer            & Yes           & No                   & Yes                & \begin{tabular}[c]{@{}c@{}}Single Paper \\ and \\ Batch of Papers\end{tabular} & Not Sure                   & No                                                                                & No                                                                   & No                                                                           & Yes                       & No                                                                            & No                                                                                                       & No                      & Yes                                                                                   \\ 
Inciteful            & Yes           & No                   & Yes                & Single Paper                                                                   & Yes                        & No                                                                                & Yes                                                                  & Yes                                                                          & Yes                       & No                                                                            & No                                                                                                       & No                      & No                                                                                    \\ 
CSAuthors            & Yes           & No                   & Yes                & Single Paper                                                                   & No                         & No                                                                                & No                                                                   & No                                                                           & No                        & Yes                                                                           & Partially                                                                                                & No                      & No                                                                                    \\
Litmaps              & Freemium      & No                   & Yes                & Single Paper                                                                   & Yes                        & No                                                                                & Yes                                                                  & Yes                                                                          & Yes                       & No                                                                            & No                                                                                                       & No                      & No                                                                                    \\ 
\textbf{InsightiGen} & \textbf{Yes}  & \textbf{Yes}         & \textbf{No}        & \textbf{Batch of Papers}                                                       & \textbf{No}                & \textbf{Yes}                                                                      & \textbf{No}                                                          & \textbf{No}                                                                  & \textbf{Yes}              & \textbf{No}                                                                   & \textbf{Yes}                                                                                             & \textbf{Yes}            & \textbf{Yes}                                                                          \\ \hline
\end{tabular}%
}
\end{table}

\clearpage
\section{CONCLUDING REMARKS}
The paper describes a tool named InsightiGen for generating meaningful and insightful information and visualization that are crucial to include in a comprehensive systematic literature review. The tool provides information including the graph-based relationship of the authors, institutions, and clusters. Moreover, it outputs trending topics of each year normalized by the previous trends, LaTex code generated for different table-based information (including influential authors and journals), adjustable sentiment analysis of each paper based on the vector area and cleansed meta-data of each paper in a JSON structured file to ease further analyses.

Future extensions of the present tool may focus on adding more insights into the citation map of the papers. The main barrier to achieving this goal is the lack of citation data in the exported data from the well-known publication repositories. With this extension, the tool will be able to calculate the paper and journal PageRank values based on the directed citation graph.

\section*{Conflict of Interest Statement}

All authors have no conflict of interest to report.

\section*{Author Contributions}

AS, and MJ, contributed to conception and design of the study. AS organized the database. AS and MJ performed the statistical analysis. AS and MJ wrote the first draft of the manuscript. All authors contributed to manuscript revision, read, and approved the submitted version.

\section*{Funding}
We would like to acknowledge the financial support of NTWIST Inc. and Natural Sciences and Engineering Research Council (NSERC) Canada under the Alliance Grant ALLRP 555220 – 20, and research collaboration of NTWIST Inc., Fraunhofer IEM, D\"{u}spohl Gmbh, and Encoway Gmbh in this research.




\bibliographystyle{Frontiers-Harvard}

\end{document}